\documentclass[aps,prb,showpacs,byrevtex,showpacs,amsmath,amssymb,twocolumn]{revtex4}
\usepackage{epsfig}
\usepackage{graphicx}

\begin{document}
\title{Local dynamical lattice instabilities: prerequisites for resonant pairing superconductivity}
\author{Julius Ranninger$^1$ and Alfonso Romano$^2$}
\affiliation{$^1$Institut N\'eel, CNRS-UJF, BP 166, 38042 Grenoble, France \\
$^2$Laboratorio Regionale SuperMat, CNR-INFM, Baronissi (Salerno), Italy, and \\
Dipartimento di Fisica ``E. R. Caianiello'', Universit\`a di
Salerno, I-84081 Baronissi (Salerno), Italy}

\begin{abstract}
Fluctuating local diamagnetic pairs of electrons, embedded in a
Fermi sea, are candidates for non-phonon-mediated superconductors
without the stringent conditions on $T_c$ which arise in
phonon-mediated BCS classical low-$T_c$ superconductors. The local
accumulations of charge, from which such diamagnetic fluctuations
originate, are  irrevocably coupled to local dynamical lattice
instabilities and form composite charge-lattice excitations of the
system. For a superconducting phase to be realized, such
excitations must be itinerant spatially phase-coherent modes. This
can be achieved by resonant pair tunneling in and out of polaronic
cation-ligand sites. Materials in which superconductivity driven
by such local lattice instability can be expected, have a $T_c$
which is controlled by the phase stiffness rather than the
amplitude of the diamagnetic pair fluctuations. Above $T_c$, a
pseudogap phase will be maintained up to a $T^*$, where this
pairing amplitude disappears. We discuss the characteristic local
charge and lattice properties which characterize this pseudogap
phase and which form the prerequisites for establishing a
phase-coherent macroscopic superconducting state.
\end{abstract}

\date{\today}
\maketitle

\section{Introduction}
Ever since the BCS theory identified the phonon exchange mechanism
as the cause for electron pairing leading to a superconducting
state and, at the same time, pinned down the order of magnitude
for the critical temperature $T_c$, considerable efforts were
undertaken to bypass these constraints on $T_c$. One of the
leading ideas was to look for fluctuating local diamagnetism
caused by intrinsic atomic correlations which would act as an
essential component for pairing in the many-body electron
wavefunction. Since diamagnetic pairs of electrons necessarily
lead to local lattice deformations, one searched for
materials\cite{Vandenberg-1977} where such deformations would
exist on a dynamical level, as close as possible to, but not at, a
global lattice instability. This was thought to favor optimal
diamagnetism and strong superconducting correlations, without
leading to localization which would originate from static lattice
instabilities.

In this line of thinking, systematic studies were undertaken
searching for compounds liable to show charge disproportionation,
originating from unstable "formal" valence states of the cations
which build up such structures. Those formal valence states,
required by the chemical stoichiometry given a certain crystalline
environment, are skipped. As a result, such materials become
composed of a mixture of cations with one more and one less
charge, respectively. Examples for such valence skippers which are
unstable are Tl$^{2+}$, Pb$^{3+}$, Sn$^{3+}$, Bi$^{4+}$,
Sb$^{4+}$. These elements prefer to exist as Tl$^{1+}$, Pb$^{2+}$,
Sn$^{2+}$, Bi$^{3+}$, Sb$^{3+}$ together with Tl$^{3+}$,
Pb$^{4+}$, Sn$^{4+}$, Bi$^{5+}$, Sb$^{5+}$. It implies a tendency
to electron pairing and for that reason has attracted, early on,
much attention for possible non-phonon-mediated superconductivity
of purely atomic origin. Charge disproportionation then originates
from effective negative-$U$ centers to which the electrons,
associated with the cations with unstable valence configurations,
are attracted and form bound electron pairs. Casting such physics
in terms of a negative-$U$ Anderson model gives rise to a variety
of macroscopic phases, such as charge fluctuation-driven
superconductivity,\cite{Ting-1980,Hirsch-1985,Schuttler-1989}
correlation-driven insulating states, translationally
symmetry-broken charge-ordered states, etc. For the exactly
half-filled band case, the stable valence configurations
(differing by two electrons) are degenerate and as a result ensure
resonant pair tunneling in and out of charge Kondo
impurities.\cite{Taraphder-1991}

The mechanism behind valence skipping has been a highly debated
subject and for a long time was attributed
\cite{Robin-1967,Chernik-1982,Moizhes-1983,Varma-1988} to an
intrinsic intra-atomic interplay between the electron affinity and
the ionization energy, combined with a crossing of stable closed
4$s$, 5$p$, 6$s$ electron shells. It has become increasingly
evident in recent years that valence skipping and the resulting
charge disproportionation systematically appear together with
distinct deformations of the cation-ligand complex, which strongly
depend on which valence state they are in. Thus, in order to
overcome the relatively large on-site Coulomb
repulsion,\cite{Matsushita-2005} strong Madelung potentials and
strong covalent bonding,\cite{Jorgensen-1989,Hase-2007} an
adequate polarizibility of the material and sufficiently large
dielectric constants ($\simeq 20-40$)\cite{Ravitch-2003} must play
together so that charge disproportionation can be realized. It is
now largely considered that the prime mechanism for charge
disproportionation lies in the local lattice relaxations of the
ligands surrounding those cations.\cite{Harrison-2006} This
picture, closer to the traditional scenario originally
hypothesized for localized defects in
semiconductors,\cite{Anderson-1975} however now requires taking
into account a time retarded - rather than static - exchange
interaction between bound electron pairs and itinerant electrons.

The question we address in this paper is to what extent charge
disproportionation can be realized in form of dynamical
diamagnetic double charge fluctuations on effective sites,
composed of cations and their surrounding ligand environments. The
latter are deformed by local lattice instabilities which change
their bond-lengths or angles. In this way, two electrons are
either captured from the Fermi sea in form of a self-trapped
bipolaron or are released onto the cations in the immediate
vicinity where they become itinerant again. One can thus hope for
a resonant tunneling of pairs of electrons in and out of such
polaronic sites, inducing local dynamical pair correlations in the
itinerant sector of the electron system. In this sense our
scenario is similar to that of fermionic atomic gases in which a
global superfluid phase is induced by a Feshbach
resonance\cite{Feshbach-1958} which describes the charge exchange
interaction between bound electron-spin triplet states and
electron-spin singlet scattering states. For such coherent pair
tunneling to occur in our counterpart solid state example, the
charge and ligand deformation fluctuations will have to be highly
correlated with each other. We shall examine here to what extent
this is feasible. If the exchange of the two types of double
charge carriers is sufficiently efficient, such charge
disproportionation materials and, possibly in a wider sense, the
copper oxides and their non-BCS like superconductivity, could be
due to the resonant bipolaron mechanism described above.

Particularly well studied examples of such materials are:

(A) Pb$_{1-x}$Tl$_x$Te\cite{Chernik-1982, Moizhes-1983} with the
parent compound PbTe which is a small gap semiconductor. Upon
doping with Tl, which exist as Tl$^{3+}$ in this compound and
plays the role of a kind of negative-$U$ center,
Pb$_{1-x}$Tl$_x$Te becomes a superconductor with a relatively high
value of $T_c \simeq 1.5\,$K, considering its small carrier
density. For details we refer the reader to an extensive
literature on this subject, see refs.\,7, 8 and references
therein.

(B) BaBi$_x$Pb$_{1-x}$O$_3$\cite{Sleight-1975} and
B$a_{1-x}$K$_x$BiO$_3$ with the parent compound BaBiO$_3$ which is
a diamagnetic insulator with a charge-ordered state involving
alternating Bi$^{\rm III}$ and Bi$^{\rm V}$ (from now on we shall
use the more appropriate denomination accounting for covalency for
the formal valence states, indicated by roman numeral in the
superscripts). Band theory predicts a metal on the basis of a
formal valence state Bi$^{\rm IV}$. The charge-ordered state is
stabilized due to a highly anisotropic polarization of the $O$
ligand environment,\cite{Hase-2007} inciting positional changes in
the two stable valence species of the Bi ions. Bi$^{\rm V}$ is in
a regular octahedra ligand environment, with a Bi$-$O distance of
$2.12 \,{\rm \AA}$. Bi$^{\rm III}$ is in a pseudo-octahedral
ligand environment where one of the six oxygen ions in the
octahedra is displaced to such an extent (with a corresponding
Bi$-$O distance of $2.28 \,{\rm \AA}$) that it is effectively
becoming O$^{2-}$ after having transferred an electron to the
cation.\cite{Simon-1988} The total outcome is a structurally
different ligand environment which favours a Bi$^{\rm III}$
valence state. Partially substituting Ba by K results in a
superconducting state with a relatively high value of $T_c \simeq
30\,$K, which has been considered as an example of
superconductivity induced by double-valence
fluctuations.\cite{Ting-1980,Hirsch-1985,Schuttler-1989}

(C) Charge disproportionation is also seen on a molecular, rather
than atomic, level in systems such as
Ti$_4$O$_7$,\cite{Lakkis-1976} V$_2$O$_5$\cite{Chakraverty-1978}
and WO$_{3-x}$,\cite{Schirmer-1980} in which one observes diatomic
pairs of Ti$^{\rm III}-$Ti$^{\rm III}$ together with
Ti$^{4+}-$Ti$^{4+}$, V$^{\rm IV}-$V$^{\rm IV}$ together with
V$^{\rm V}-$V$^{\rm V}$, and W$^{\rm V}-$W$^{\rm V}$ together with
W$^{\rm VI}-$W$^{\rm VI}$, respectively. These different valence
states of the molecular units again show distinctly different
intra-molecular distances, such as $2.8 \,{\rm \AA}$ for Ti$^{\rm
III}-$Ti$^{\rm III}$ and $3.08 \,{\rm \AA}$ for Ti$^{\rm
IV}-$Ti$^{\rm IV}$ bonds. Moreover, with increasing temperature
these systems exhibit a thermally activated conductivity by pair
hopping. This demonstrates the stability of such bound electron
pairs on deformed molecular units which are locked together in a
dynamical fashion with the charge fluctuations.

(D) Indications for dynamical charge fluctuations are also seen in
the cuprate superconductors. Here the situation is however more
involved. Stable valence states of Cu exist as:

(i) Cu$^{\rm I}$ in a dumbbell oxygen ligand environment, with a
characteristic Cu$-$O bond-length of $1.84 \,{\rm \AA}$ (examples
are Cu$_2$O, BaCu$_2$O$_2$, FeCuO$_2$);

(ii) Cu$^{\rm II}$ in square planar oxygen ligand environments
with a characteristic  Cu$-$O bond-length of $1.94 \,{\rm \AA}$
(an example is CuO$_4$);

(iii) Cu$^{\rm III}$ in square planar oxygen ligand environments,
in rare cases, with a characteristic Cu$-$O bond-length of around
$1.84 \,{\rm \AA}$, and hence practically identical to that of
Cu$^{\rm I}$ in dumbbell environments\cite{Simon-1988} (an example
is KCuO$_2$).

In the cuprates the formal valency of Cu lies between $\rm II$ and
$\rm III$. The square lattice of the cuprates is neither a strict
square planar nor a dumbbell ligand environment and there is no
indication for a static disproportionation of Cu$^{\rm II}$.
Cu$^{\rm III}$, which should be favored in this cuprate planar
lattice structure, could appear only through rapid metallic
fluctuations or be due to the stoichiometry of the structure, with
an atypical Madelung potential.\cite{Wilson-1987} Yet, local
dynamical lattice fluctuations involving the oxygen ligands of the
planar Cu ions, the Cu$-$O bond stretch (or buckling) modes,
exist. They can be considered as being linked to charge
fluctuations involving local ligand environments which are typical
of stable Cu$^{\rm I}$ and Cu$^{\rm III}$. In the anti-nodal
region of the Brillouin zone, where the electronic density of
states (DOS) is dominated by the the pseudogap phenomenon, local
scanning tunneling microscope (STM) spectroscopy\cite{Lee-2006}
did establish a strong coupling of electrons with wave vectors
$[k_x, k_y] = [\pm \pi,0]$ and $[0, \pm \pi]$ and local phonon
modes with characteristic frequencies distributed around
$50\,$meV. The isotope variation of these modes (most likely
associated to the Cu$-$O bond stretch mode) correlates with that
of the pairing gap. This strongly indicates that this local Cu$-$O
bond stretch mode could play a significant role in the pairing
mechanism of the cuprates.

If double charge fluctuations do play a significant role in
establishing a superconducting phase, they have to sustain a free
particle-like behavior with spatial phase coherence. This should
occur in spite of the sizeable lattice relaxation which, a priori,
tends to render them rather diffusive. If in this largely
diffusive dynamics we can obtain a finite component of coherent
phase-correlated double charge fluctuations, then a
superconducting state can materialize. Such a state does not
describe a purely bosonic system of bipolarons and bipolaronic
superconductivity, but is characterized by resonating bipolarons
embedded in a Fermi sea. Their signature is a pseudogap in the
single-particle DOS at the Fermi level due to strong pairing
correlations above the superconducting phase. This scenario has
been discussed in great detail over the past years on the basis of
the Boson-Fermion Model by J. Ranninger and collaborators.

To address the question of the feasibility of coherent pair
tunneling, we consider charge carriers capable of existing either as
selftrapped bipolarons or as itinerant quasi-free tight-binding
charge carriers on a small clusters consisting of:

(i) cation-ligand complexes which are deformable and which can
capture the charge carriers  in form of bipolarons;

(ii) square plaquettes composed of four structurally identical
cations-ligand complexes (surrounding the deformable central
cation-ligand complex), which we consider as undeformable. Charge
carriers on these plaquettes shall hence move as itinerant
entities rather than being captured by polaronic effects.

Double charge fluctuations in the present picture implies
fluctuations between doubly occupied and unoccupied sites, driven
by charge carriers tunneling in and out of such sites and
ultimately resulting in pair correlations among those on the
plaquette sites. Such essentially local physics contains the
essence of triggering a crossover from an insulating purely
charge-disproportionated state to a superconducting state upon
tuning certain parameters such as the electron-lattice coupling,
the adiabaticity ratio or the charge carrier density. This
physics, intrinsically related to a very local
mechanism,\cite{Ranninger-2006} manifests itself in the single
particle spectral function, the local diamagnetism, the local
phonon softening, the quasi-elastic peak in the neutron scattering
cross-section for the phonons and the double-peaked pair
distribution function for the bond-length fluctuations.

In section II we discuss the salient features of such double
charge fluctuations driven by bond-length fluctuations and present
a scenario in terms of a kind of Holstein model which takes into
account the different ligand fluctuations characterizing the
various charge configurations. In section III we propose a model
Hamiltonian which we consider to be adequate to describe the local
physics of such systems. We discuss its basic features as far as
the efficiency of local pair tunneling is concerned and how such
dynamically correlated charge-ligand deformation fluctuations are
locked together in order to form phase-coherent correlated
itinerant excitations on a short spatial scale. In section IV we
discuss the specific characteristics of: (i) the local lattice
dynamics close to instabilities, phonon softening, the
quasi-elastic peak in the dynamical structure factor and the
double peak structured pair distribution function,  (ii) the
diamagnetic correlations induced among the itinerant electrons via
such dynamical local lattice instabilities and  how they manifest
themselves in a pseudogap feature in the electronic density of
states. In the Summary, section V, we briefly review the main
findings of the present work and discuss the further issues which
have to be treated next.

\section{Electron localization versus delocalization in electron-lattice coupled systems}

The crossover from metallic, or superconducting, to localized
behaviour in electron-lattice coupled systems has been a topic of
longstanding interest and debate.\cite{Varenna-2006} The issue has
been reinvestigated with renewed vigor in recent years, when it
became apparent that an insulating phase of localized polaronic
charge carriers, bordering on a superfluid phase, can be induced
by many-body correlation
effects.\cite{Benedetti-1998,Meyer-2002,Capone-2003} Earlier
widely pursued ideas considered density-driven stripping off of
the phonon clouds of individual
polarons\cite{Hohenadler-2003,Hohenadler-2005} and a related to it
possible breakdown of Wigner polaron
crystallization\cite{Quemerais-1995,Quemerais-2004} into a
metallic phase. Correlation-driven localization can be turned into
correlation-driven superconductivity by small modification of
certain tunable experimental parameters. Real systems where such
and similar phase changes are manifest (cuprates, manganites,
transition metal oxides, magnesium diborides, just to name a few)
are complex in their structure as well as in the composition of
their metal-ion-ligand electronic configurations. Their
particularity lies in a prevailing local physics, which has been
generally overlooked in theoretical studies of the polaron
problem. This local physics nevertheless dictates the outcome of
macroscopic states such as (i) a superconducting phase of
itinerant electrons with Cooper pairing, (ii) an insulating phase
of localized bipolarons, and, possibly, (iii) a phase
corresponding to a Bose metal sandwiched between the two, which
has become of particular interest in recent
years.\cite{Phillips-2003,Stauber-2007,Cuoco-2006} Since the physics of
electron-lattice coupled systems involves local atomic or
molecular displacements, it requires a description in terms of
small polarons, i.e., electrons dragging with them a sizable local
lattice deformation which can be of the order of a tenth or so of
the lattice constant. We have for that reason to rely on the
picture of a Holstein-type molecular crystal model as the basis of
our study.

As a paradigm for systems with resonant pairing induced by local
lattice deformations, we can imagine as the simplest scenario a
bipartite lattice like the one illustrated in Fig.\,1 and
consisting of:

(A) a sublattice of polaronic sites, where the charge carriers
couple locally and strongly to the molecular deformations of the
ligands, resulting in localized bipolaronic bound pairs in the
cation-ligand bonds;

(B) a sublattice of non-polaronic sites on which the electrons
(holes) move as itinerant charge carriers.

\begin{figure}[tbp]
 \begin{center}
    \includegraphics*[width=3in,angle=0]{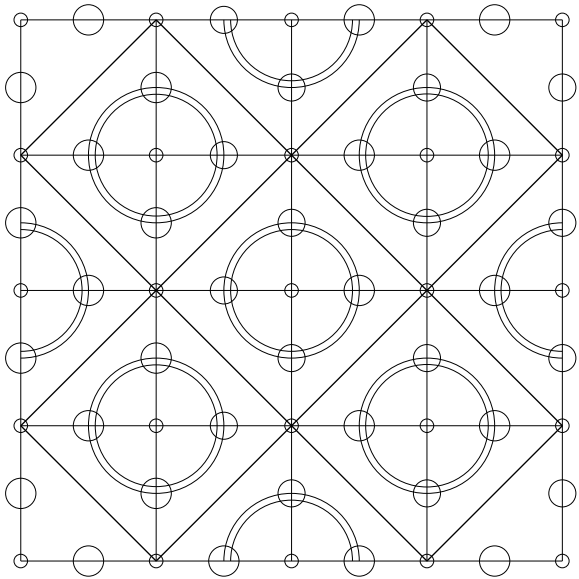}
\caption{\footnotesize Paradigm of a 2D resonant pairing system in
terms of a bipartite lattice composed of polaronic cation-ligand
complexes (rings) comprising four anions (large circles) in the
center of interlinked plaquettes of four metallic cation sites
(small circles) housing the charge carriers in itinerant states.}
\end{center}
\end{figure}

When the charge transfer between the two sublattices is switched
on, the electron (hole) pairing on polaronic sites, leading to
localized bipolarons, will compete energetically with the
itinerancy of uncorrelated charge carriers on the non-polaronic
sites in their immediate vicinity, when their respective energies
are comparable. This induces pair correlations among the otherwise
intrinsically uncorrelated itinerant charge carriers in locally
very {\bf confined} regions of the lattice. On a macroscopic
scale, this local physics (when treated on the basis of a
heuristic Boson-Fermion Model scenario) ultimately results in a
competitions between a correlation-driven insulating phase and a
superconducting one.\cite{Cuoco-2004,Cuoco-2006} The transition
from one phase into the other can be  tuned  by varying the
strength of the electron-lattice coupling, the size of the
adiabaticity ratio, the relative ionic potential difference
between the polaronic and the non-polaronic sites, and the
concentration of the charge carriers. Specific interconnected
local properties in such systems, which are determinant for the
physics on a macroscopic scale, are:

i) the generation of an intrinsic pseudogap feature in the local
electronic density of states\cite{Cuoco-2006,Stauber-2007} which,
depending on the parameters of the system, foreshadows an
insulating or a superconducting gap;

ii) the  strong softening and a concomitant increase of spectral
weight of local vibrational modes on the polaronic
sites,\cite{Ranninger-2006} serving as a basic ingredient for the
bipolarons to acquire itinerancy and assuring a component of
bosonic charge carrier transport;

(iii) the double-peak structure in the local lattice displacement
pair distribution function due to a quantum coherent
superposition\cite{Ranninger-2006} of bipolarons and pairs of
intrinsically uncorrelated electrons in their immediate vicinity,
which can result in dynamically fluctuating stripe-type
topological structures.

The approach to the polaron problem followed here is similar to
procedures followed in the correlation problem,\cite{Altman-2002}
based on Anderson's Resonating Valence Bond
idea.\cite{Anderson-1987}. There, an intrinsic underlying magnetic
structure leads to the formation of local singlet hole pairs on
plaquettes, dynamically exchanging with pairs of uncorrelated
itinerant holes in their immediate vicinity. As concerns the
connections with other electron-phonon models, the present
paradigm of a fluctuating bipartite lattice structure of polaronic
and non-polaronic effective sites is qualitatively different from
the usually studied molecular crystal Holstein model, where all
the sites are considered to be equivalent. Moreover, when the
attention  is focused on charge Kondo-like resonance
features,\cite{Capone-2006,Hewson-2002,Koller-2004} the local
electron-phonon coupling is assumed to be of the form $\hbar
\omega_0 \alpha (n_{i,\uparrow} + n_{i,\downarrow} -1)[a^+_i
+a_i]$, i.e. it is at any site symmetric with respect to the
single-site occupation. On the other hand, as we have stressed in
the beginning, systems with double charge fluctuations show strong
anisotropy of their ligand deformations. This motivates us to take
an asymmetric electron-lattice coupling of the form $\hbar
\omega_0 \alpha (n_{i,\uparrow} + n_{i,\downarrow})[a^+_i +a_i]$.
Although the modulus of the displacements of either unoccupied or
doubly occupied sites relative to the singly occupied sites, i.e.
$\langle a^+_i + a_i \rangle = \pm 2\alpha(n_{i,\uparrow} +
n_{i,\downarrow})$, is the same for the two types of couplings
(symmetric and asymmetric), the respective energies, related to
the electron-lattice coupling, are not. For the usual symmetric
electron-phonon coupling the energies of the unoccupied and doubly
occupied site are degenerate. It is this which results in the
Kondo-type resonance in the charge
channel,\cite{Hewson-2002,Koller-2004,Capone-2006} provided that
the chemical potential of the system is such that this degeneracy,
representing the two stable valence configurations, is guaranteed,
which is the case for the exactly half-filled band limit. Away
from this limit this is not the case and the resonance
disappears.\cite{Koller-2004}. A further and essential difference
between the generic Holstein scenario and the present paradigm for
resonant pairing systems, is that in the valence fluctuation
scenario the ligands of two adjacent cations have an anion in
common. This leads to strong intersite correlations between
cations with different valence states and correspondingly
different ligand environments compatible with this bridging
oxygen. These facts differentiate between sites where the
bipolarons form and sites in the immediate vicinity where they do
not. This is similar to what happens with Zhang-Rice singlets in
cuprates,\cite{Hirsch-1987,Zhang-1988} where singlet pairs form on
a Cu ion and its immediate ligand environment of four oxygens, but
where nearest-neighboring Cu ions cannot behave as such Zhang-Rice
singlets at the same time. This "exclusion"
feature\cite{Roehler-2005} introduces in the end strong local
correlation effects in the band structure of the whole system,
which is part of the salient features of the cuprates. Our
proposition here is that the local polaronic features of double
charge fluctuations should have a similar effect on the overall
band structure and in particular will give rise to a charge
pseudogap of polaronic origin.

\section{The model}
The aim of this work is to examine the local physics of polaronic
systems lending themselves to double charge fluctuations, coupled
via an exchange term to itinerant charge carriers. They impose on
the latter strong local dynamical diamagnetic fluctuations which
can eventually lead to a global phase-locked superconducting
state. We shall do this here, having in mind the cuprates and in
particular their normal phase properties. These systems are more
complicated than the paradigm for resonant pairing systems
discussed in the previous Section, in the sense that all the Cu
cations are intrinsically equivalent and there is no translational
symmetry breaking to be expected from the outset. Yet, we had
early-on strong experimental indications\cite{Toby-1990} that the
structural correlations and deformations in the cuprates show
remnant bipartite lattice features, on a local level as well as on
a finite time scale. These topologically different features,
generally referred to as "stripes", break the translational, as
well as the rotational, symmetry on a local level.

The application of the valence fluctuation scenario, describing
possible resonant pairing in high-$T_c$ superconducting cuprates,
implies, for the reasons clarified below, the consideration of
three formal valence configurations in the CuO$_2$ layers,
Cu$^{\rm I}$, Cu$^{\rm II}$ and Cu$^{\rm III}$. Distinct ligand
deformations are associated with them, dumb-bell for Cu$^{\rm I}$
and square planar for Cu$^{\rm II}$ and Cu$^{\rm III}$, but
differing from each other by the value of the Cu$-$O distances
(see Fig.\,2). The Cu cations in the CuO$_2$ layers of the
insulating parent compound have a square oxygen ligand environment
with Cu$-$O distances equal to 1.94 $\rm \AA$ and are in formal
valence states Cu$^{\rm II}$, according to their chemical
composition. Moreover, the cuprates represent a metastable
structure characterized by a geometric misfit of the various
layers\cite{Sleight-1991} which leads to Cu$-$Cu distances in the
CuO$_2$ layers which can accommodate such [Cu$^{\rm II}-$O$_4]$
units (see Fig.\,2a) only by pushing the oxygens alternately
slightly above and below these layers. This structural effect is
usually called buckling.

Upon hole doping one introduces single holes in such [Cu$^{\rm
II}-$O$_4]$ units, which make them turn into [Cu$^{\rm
III}-$O$_4]$ units, with shortened Cu$-$O distances equal to
1.84$\,\rm\AA$. Referring to our small cluster, such a doped
effective cation-ligand unit is represented by the central
effective site 5 in Fig.\,2b. The reduction to 1.84$\,\rm\AA$ of
the bond-lengths in the [Cu$^{\rm III}-$O$_4]$ units is
accompanied by an unbuckling of the four oxygen cations belonging
to these units, which are pulled into the CuO$_2$ plane, but keep
unaltered their distances from the Cu$^{\rm II}$ ions on the
plaquette sites, thus inducing no long-range stress field. In this
way, the oxygen cations surrounding each of the Cu$^{\rm II}$ ions
(located at sites 1-4 in Fig.\,2b) are still all at the same
distance 1.94$\,\rm\AA$ from it, but one of them [the bridging
oxygen between Cu$_5$ and Cu$_i$ ($i$=1$-$4)] now lies in the
CuO$_2$ plane and not above or below it.

We conjecture that the [Cu$^{\rm III}-$O$_4]$ units with their
doping-induced ligand deformations act as polaronic traps, binding
two holes in form of a localized bipolaron. If the energy of this
state is almost degenerate with that of the hole states at the
Fermi level, then this bipolaron can split into two holes going on
the surrounding matrix, and, as a consequence, [Cu$^{\rm
III}-$O$_4]$ turns into [Cu$^{\rm I}-$O$_4]$. As pointed out in
the previous Section, Cu$^{\rm I}$ is typically in a linear
O$-$Cu$-$O dumb-bell configuration. If we suppose that this
happens also in the CuO$_2$ layers, we are faced with two
degenerate O$-$Cu$-$O dumbbell bonds along the $x$ and $y$
directions, with a shortened bond-length of 1.84$\,\rm \AA$ (see
Fig.\,2c) which is the same as for the [Cu$^{\rm III}-$O$_4]$
units. This would cause an anisotropic unbuckling of the
corresponding bridging oxygens in either one or the other
direction, but leaving the Cu$^{\rm II}-$O$_4$ units on the
plaquette sites unaltered, as far as their Cu$-$O bond-lengths are
concerned. It is also feasible that, given this directional
degeneracy of the bonds of the Cu$^{\rm I}$ valence states, an
isotropic ligand deformation with the same bond-length of
1.84$\,\rm\AA$ materializes, due to $d_{x^2 -y^2}$ hybridization
in the CuO$_2$ plane\cite{Roehler-2008} and which then would be
very similar to the configuration illustrated for the Cu$^{\rm
III}-$O$_4$ unit in Fig.\,2b. It would lead to an isotropic
unbuckling, again with no changes in the bond-lengths of the
cation-ligand units on the plaquettes.

The microscopic foundation of the various formal valence
configurations, conjectured here to play a role in the cuprates,
are left to be investigated in the future. For the present study,
where we shall treat the ligands as an effective deformable scalar
quantity, we do not deal with such detailed questions concerning
the anisotropy of the deformations on the central sites acting as
resonating bipolaronic traps. The valence fluctuations involving
the central [Cu$^{\rm III}-$O$_4]_5$ unit lead to the release of
two holes from the latter onto the neighboring [Cu$^{\rm
II}-$O$_4]_i$ units on the plaquette sites $i$=1$-$4, and
vice-versa ([Cu$^{\rm III}-$O$_4]_5 \Leftrightarrow $[Cu$^{\rm
I}-$O$_4]_5$ + two holes). Once on the plaquette units, these
holes engage in itinerant states which prevent them from getting
localized in form of self-trapped bipolarons on those [Cu$^{\rm
II}-$O$_4]_i$ units. Such a mechanism describes tunnelling in and
out of hole pairs from a given central deformable ligand
[Cu$-$O$_4]_5$ unit (our effective site 5) into the surrounding
[Cu$^{\rm II}-$O$_4]_i$ units which remain essentially undeformed.
For this reason we neglect here any charge-ligand deformation
coupling on those plaquette sites.

We remind the reader that formal valence in such systems has to be
interpreted in terms of the charges involved in the covalent bonds
between the cations and their ligand environments and not simply
by the charges on the cation. In the cuprates the covalent
character in the CuO$_2$ planes is very pronounced. This might be
one reason why spectroscopically neither Cu$^{\rm III}$ nor
Cu$^{\rm I}$ have been detected sofar. Furthermore, given the fact
that the three formal valence states Cu$^{\rm I}$, Cu$^{\rm II}$
and Cu$^{\rm III}$ have different ligand environments, the
coupling of those deformations to the charge will in general not
be symmetric with respect to singly occupied units [Cu$^{\rm
II}-$O$_4$]. In the planar CuO$_2$ layer of the cuprates all the
Cu ions are equivalent and their charge coupling to the ligand
deformations are spatially homogeneous.  But this coupling does
depend on the formal Cu valence state which is active at any given
moment. If we consider the cuprates as being made out of an
ensemble of overlapping clusters, as given in Fig.\,2, this
immediately leads to local dynamical correlations between
neighboring ligand deformations. On a short range we can therefore
expect that this will naturally lead to intercalated topological
local structures. The anisotropy of the O$-$Cu$_{\rm I}-$O bonds
could be relevant for the anisotropic local symmetry breaking.

\begin{figure}[tbp]
\begin{center}
   \includegraphics*[width=1.8in,angle=0]{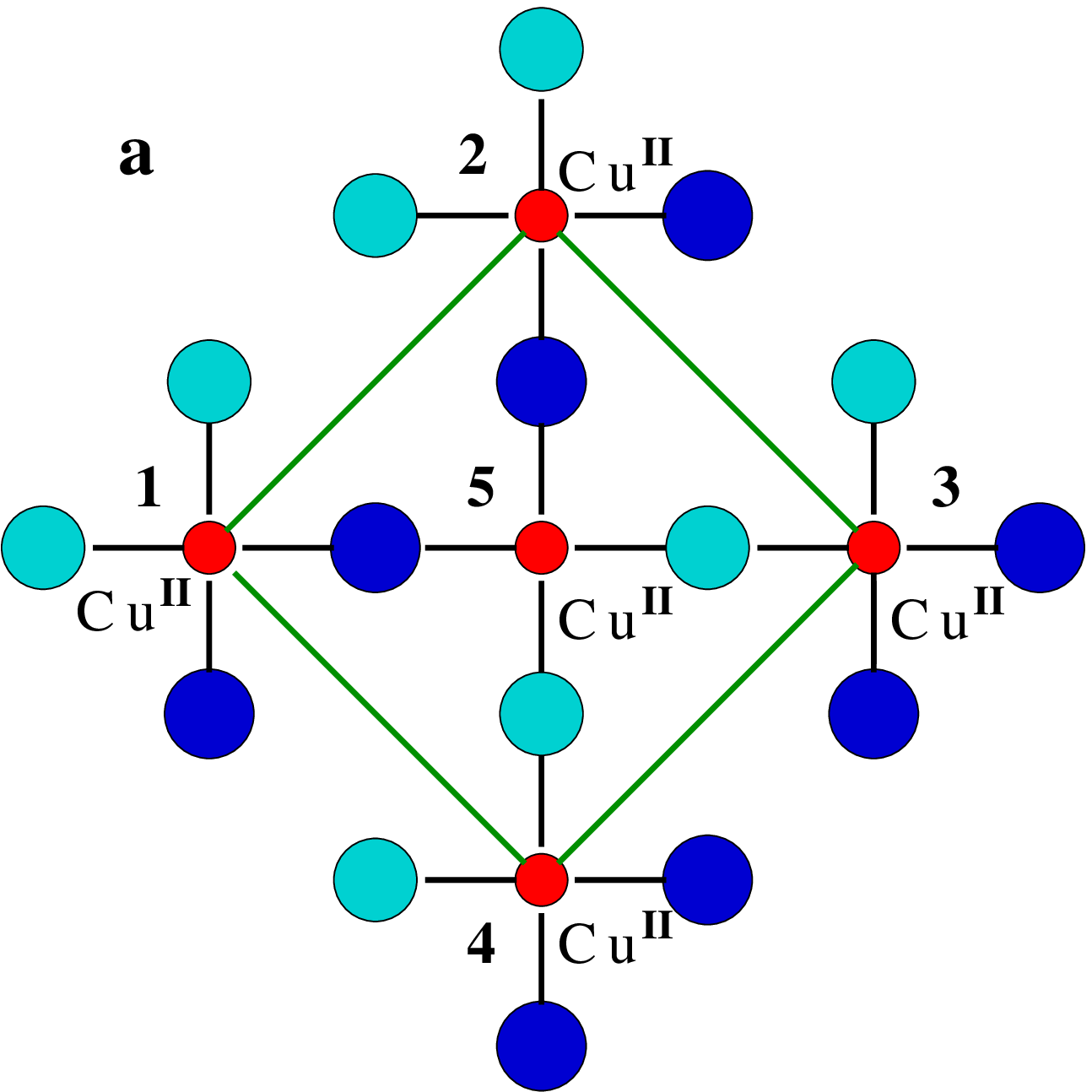}
\end{center}
\begin{center}
   \includegraphics*[width=1.8in,angle=0]{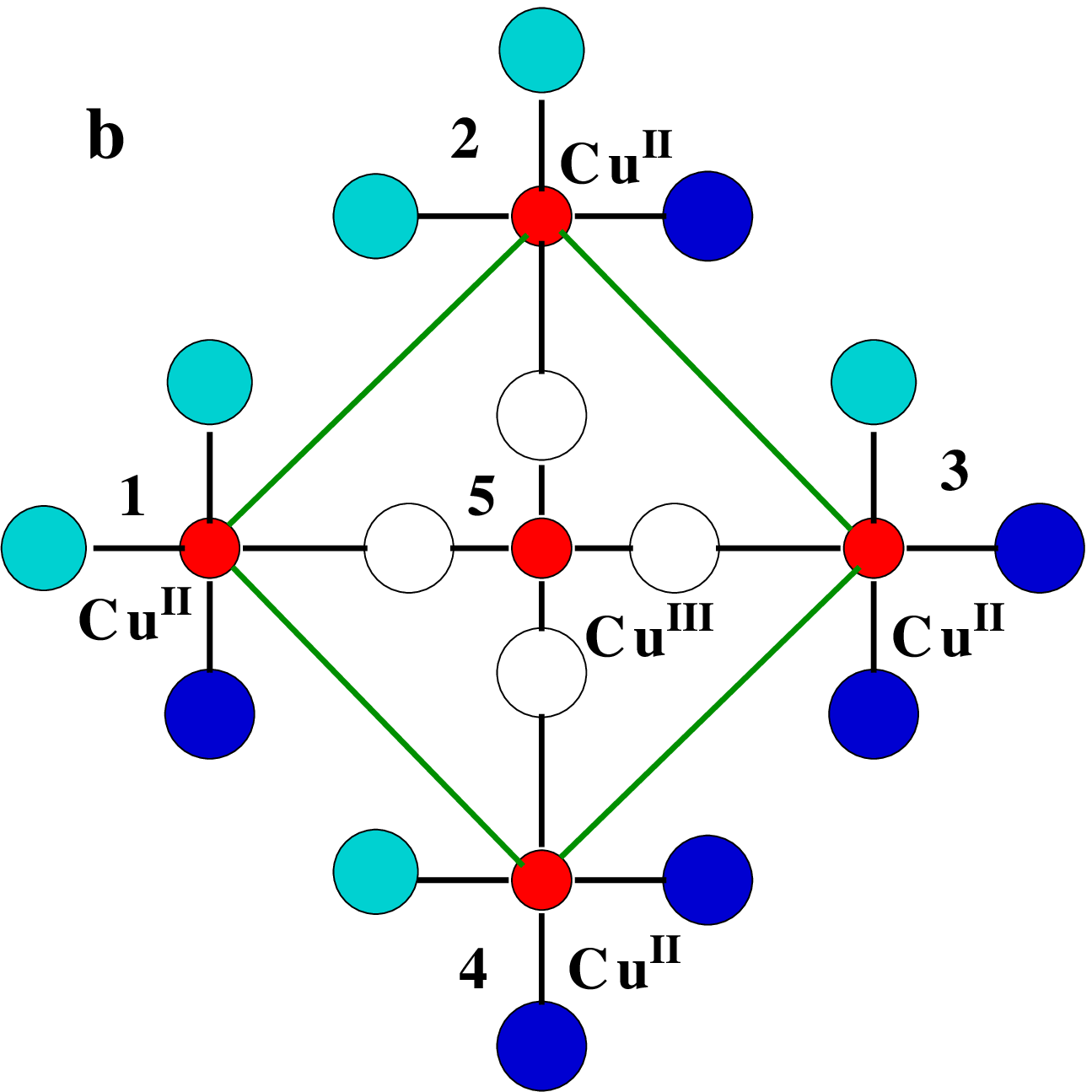}
\end{center}
\begin{center}
   \includegraphics*[width=1.8in,angle=0]{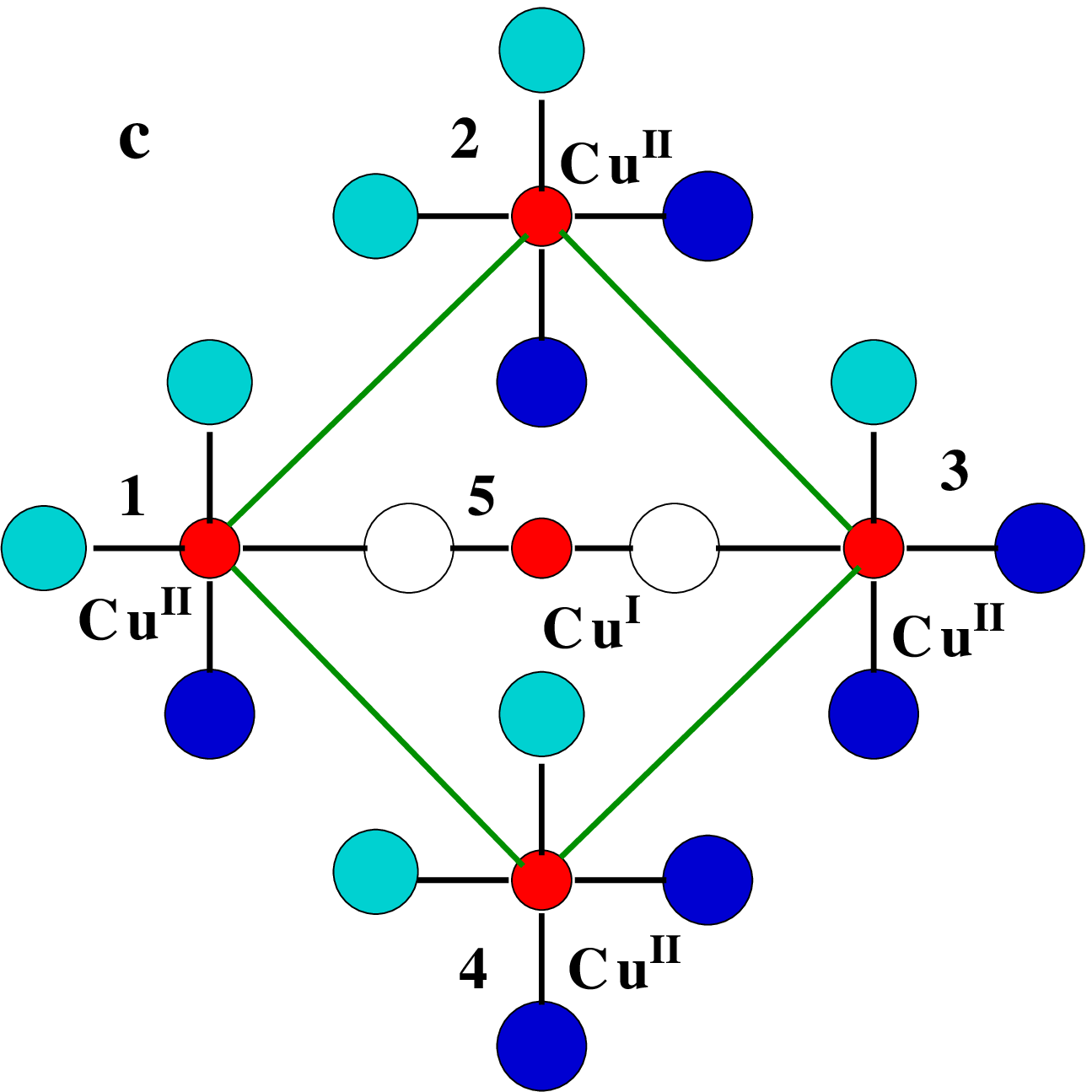}
\caption{\footnotesize (Color online) Basic cluster in double
valence fluctuation-driven pairing in the CuO$_2$ planes
consisting a central Cu cation oxygen ligand environment in three
different formal valence states, Cu$^{\rm II}$, Cu$^{\rm III}$ and
Cu$^{\rm I}$, illustrated in Figs.\,2a, 2b and 2c, respectively.
Such central cation-ligand units are linked via a common bridging
oxygen to neighboring Cu$^{\rm II}$-O$_4$ units. The small red
circles denote the Cu cations and the large circles the oxygens.
Dark blue means that they are pushed above the CuO$_2$ planes,
light blue that they are below and white circles that they are
in-plane.}
\end{center}
\end{figure}

The construction of a macroscopic state on an infinite lattice,
starting from a spatially translationally invariant Hamiltonian
but working in a representation of such finite cluster states,
which contain the essential local physics, can in principle be
achieved. One way would be using  renormalization group techniques
such as the plaquette contractor method, based on small clusters
of the form illustrated in Fig.\,2. This method was developed by
Morningstar and Weinstein\cite{Morningstar-1996} and applied to
the RVB scenario by Altman and Auerbach\cite{Altman-2002}. For the
present scenario of valence fluctuation driven pairing, including
the various ligand deformations correlated to the formal valence
state of the clusters, presents a formidable numerical task.
Questions like that will be addressed in some future work.

Here we shall content ourselves to study the dynamics of such
basic clusters making up the CuO$_2$ layers, which permits us to
draw some preliminary conclusions on such macroscopic phases.
Considering an ensemble of such spatially and temporarily
uncorrelated fluctuating clusters, without any long-range spatial
phase coherence between them, allows us to describe certain
features of the pseudogap phase in the cuprates and its
corresponding onset temperature $T^*$. We can assess the frequency
of the resonant pair tunneling processes, driven by the local
ligand deformation fluctuations (which play the role of local
dynamical lattice instabilities in a macroscopic system) and which
can be approximately related to the mass of the diamagnetic pair
fluctuations and thus to the value of $T_c$ for the onset of a
Bose-Einstein condensation-driven superconductivity.

In order to capture the salient features of the local physics of
valence fluctuations in the cuprates, as exposed above, we
illustrate in Figs.\,2a-c. the various valence configurations of
such clusters, where the central Cu cation is alternatively in a
Cu$^{\rm I}$, Cu$^{\rm II}$ and Cu$^{\rm III}$ formal valence
state. We address this local problem here in a somewhat simplified
version, neglecting  the detailed structure of the oxygen ligand
environments surrounding the Cu cations. For that purpose we
assume the [Cu$-$O$_4]_5$ unit to be composed of (i) a central
cation surrounded by an isotropically deformable ligand
environment of four anions on an effective site 5, and (ii) four
neighboring [Cu$-$O$_4]_i$ units ($i$=1-4) on the plaquette,
representing non-polaronic sites for which we consider the
charge-ligand deformation coupling to be inactive. This is because
we consider the charge carriers on the plaquette sites to move as
itinerant entities and those ligands have no time to respond to
the charge-deformation coupling. They remain in their buckled
configurations, corresponding to the [Cu$^{\rm II}-$O$_4]_i$
units. We shall denote the hopping rate between the plaquette
sites by $t$ and that between the latter and the central polaronic
deformable unit by $t^*$. The dimensionless coupling between the
charge density and the local ligand deformations is denoted by
$\alpha$ and the bare vibrational frequency of the bond-length
fluctuations of that latter by $\omega_0$. We assume that the
ionic levels of the plaquette sites and of the polaronic sites
have a difference in energy, given by $\Delta$. The reason for
that is the following. Our picture is that the local clusters, as
illustrated in Fig.\,2, represent the essential dynamical units of
the cuprate CuO$_2$ planes. In this picture, the states of holes
on the plaquette sites play the role of the states of itinerant
holes in an infinite system close to the Fermi level, moving in a
Hubbard correlated fashion. The holes we tackle are those induced
by doping the parent compound of the correlated problem but which,
on an individual cluster, we shall simply treat as itinerant
uncorrelated holes on the plaquette. Including the Hubbard $U$ in
this problem would require one to consider at the same time the
electronic correlations on the polaronic sites, leading to
Zhang-Rice singlets, and those on the plaquette sites, reflecting
their pairing in the overall density of states near the Fermi
level. This is, however, beyond the present description.

Given these considerations, the Hamiltonian describing such
cluster is
\begin{eqnarray}
H & = & - \, t \sum_{i \ne j = 1 ... 4, \,\sigma} \left[
c^{\dagger}_{i,\sigma} c_{j,\sigma}
\, + \, h.c.  \right]  \nonumber \\
& -& \, t^* \sum_{i=1 ... 4, \,\sigma} \left[
c^{\dagger}_{i,\sigma} c_{5,\sigma} \, + \, h.c.  \right] \, + \,
\Delta \, \sum_{\sigma}
c^{\dagger}_{5,\sigma} c_{5,\sigma}  \nonumber \\
& + &\, \hbar\omega_0 \left[a^{\dagger} a + \frac{1}{2}\right] \,
- \hbar\omega_0\alpha \, \sum_{\sigma} \, c^{\dagger}_{5,\sigma}
c_{5,\sigma}\, \left[a + a^{\dagger}\right], \; .
\end{eqnarray}
where $c^{(\dagger)}_{i\sigma}$ denote the annihilation (creation)
operator for a hole with spin $\sigma$ on site $i$, and
$a^{(\dagger)}_5$ the phonon annihilation (creation) operator
associated with the deformable cation-ligand complex at site $5$.
This local Hamiltonian describes a competition between localized
bipolaronic hole pairs on the central cation-ligand complex and
itinerant holes on the plaquette sites, when the energies of the
two configurations are comparable, i.e. for $2\Delta - 4 \hbar
\omega_0 \alpha^2  \simeq -4t$. We illustrate in Fig.\,3 the
variation with $\alpha$ of the low energy spectrum of such a
system for a set of parameters ($t=0.2$, $t^* = 0.15$, $\Delta =
0.5$ and $\omega_0 = 0.1$, in bare units) corresponding to a
typical region in the parameter space $[\omega_0/t, \hbar
\omega_0\alpha^2 /t]$ where resonant pairing is well pronounced
and rather robust against thermal fluctuations. Our approach is a
straightforward exact diagonalization of $H$ leading to
eigenenergies $E_n$ associated with eigenstates of the form
\begin{equation}
|n\rangle = \sum_k\sum_{\nu=0}^{N_{\rm ph}}
A^{(n)}_{k,\nu}\,|k\rangle \otimes |\nu \rangle
\end{equation}
where $|k\rangle$ designates the configurations of two holes with
opposite spin distributed over the five cluster sites,
$|\nu\rangle$ denotes the $\nu$-th excited state of the
undisplaced harmonic oscillator state, and $A^{(n)}_{k,\nu}$ is
the corresponding weight in the eigenstate $|n\rangle$. Taking
into account a truncated Hilbert space for the phonon states
$|\nu\rangle$, we limit ourselves to $N_{\rm ph} = 70$. For the
set of parameters and temperatures we are concerned with, this is
sufficient. We notice from Fig.\,3 that the low energy eigenvalues
$E_n \simeq -4t + n \hbar \omega_0$ are relatively unaffected as
$\alpha$ is increased, except when they get close to crossing the
energy of the localized bipolaron level at some $\alpha_c$. As
$\alpha$ is increased above this value, the energy eigenvalues
follow the downward trend of the localized bipolaron energy,
tending to $E_n \simeq 2\Delta - 4 \hbar \omega_0 \alpha^2  + n
\hbar \omega_0$.
\begin{figure}[tbp]
 \begin{center}
    \includegraphics*[width=3in,angle=0]{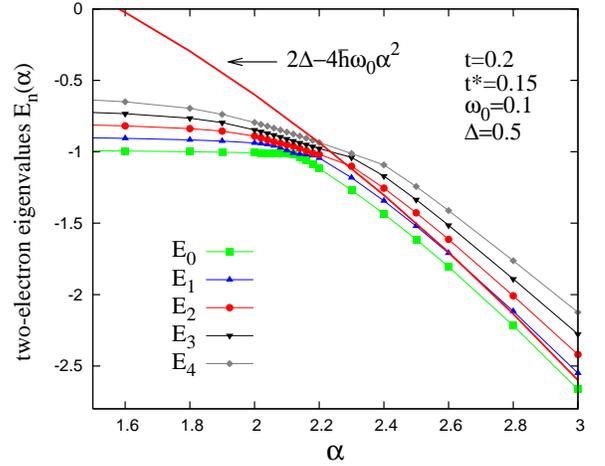}
\caption{\footnotesize (Color online) The low energy two-hole
spectrum of $H$ as a function of the electron-lattice coupling
constant $\alpha$, showing the crossover between delocalized
itinerant states on the plaquette (for $\alpha \leq
\alpha_c=2.12$) and localized bipolaron states on the central
polaronic site (for $\alpha \geq \alpha_c$).}
\end{center}
\end{figure}
\begin{figure}[tbp]
 \begin{center}
    \includegraphics*[width=3in,angle=0]{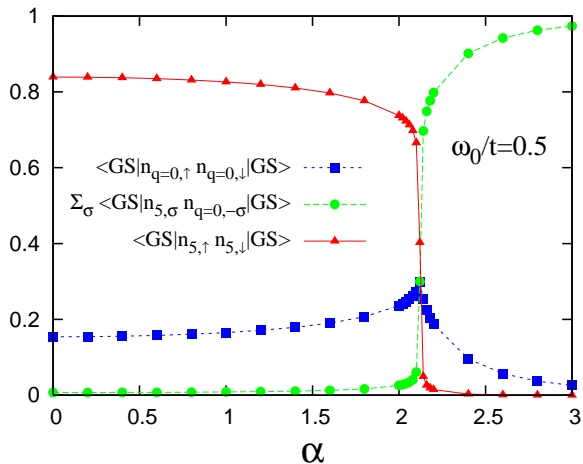}
\caption{\footnotesize (Color online) The occupation probabilities
for the various possible hole configurations as a function of
$\alpha$ and for $\omega_0/t = 0.5$.}
\end{center}
\end{figure}
\begin{figure}[tbp]
 \begin{center}
    \includegraphics*[width=3in,angle=0]{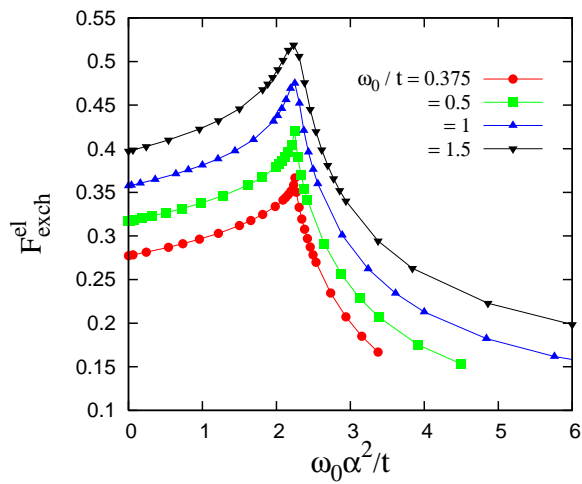}
\caption{\footnotesize  (Color online) The efficiency factor
$F^{\rm el}_{\rm exch}$ as a function of $\omega_0 \alpha^2/t$ for
a hole tunneling in and out of a polaronic site, favoring double
charge fluctuations, for several adiabaticity ratios $\omega_0/t$.
For clarity of presentation, the curves obtained for
$\omega_0/t=0.5, 1, 1.5$ have been rigidly shifted upwards with
respect to the one referring to $\omega_0/t=0.375$ by amounts
equal to 0.04, 0.08, 0.12, respectively. The numerical values on
the $y$-axis thus refer specifically to the case
$\omega_0/t=0.375$.}
\end{center}
\end{figure}
\begin{figure}[tbp]
 \begin{center}
    \includegraphics*[width=3in,angle=0]{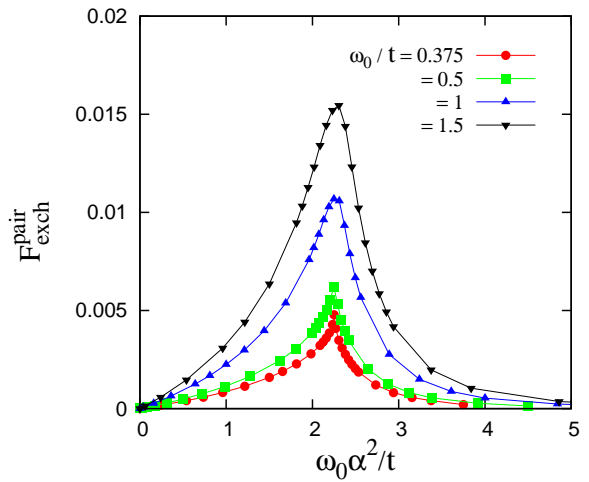}
\caption{\footnotesize  (Color online) The efficiency factor
$F^{\rm pair}_{\rm exch}$ as a function of $\omega_0 \alpha^2/t$
for converting, via resonant tunneling, a localized bipolaron on
the central polaronic cation-ligand complex into a pair of
correlated holes on the ligand environment, for several
adiabaticity ratios $\omega_0/t$.}
\end{center}
\end{figure}

In Fig.\,4 we trace out the occupation probabilities in the
two-particle ground state $|GS\rangle$ for the configurations with
(i) the two holes on the polaronic site, $\langle
GS|c^{\dagger}_{5,\uparrow}c_{5,\uparrow}c^{\dagger}_{5,\downarrow}c_{5,\downarrow}|
GS \rangle$, (ii) the two holes on the plaquette sites, $\langle
GS| c^{\dagger}_{q=0,\uparrow}c_{q=0,\uparrow}
c^{\dagger}_{q=0,\downarrow} c_{q=0,\downarrow}|GS\rangle$, where
$c^{\dagger}_{q=0,\sigma}| 0 \rangle = \frac{1}{2}\sum_{i=1}^{4}
c^{\dagger}_{i,\sigma}|0\rangle$ signifies the lowest-energy Bloch
state of the plaquette, and (iii) the two holes distributed
between the two types of sites,
$\sum_{\sigma=\uparrow,\downarrow}\langle GS|
c^{\dagger}_{5,\sigma}c_{5,\sigma} c^{\dagger}_{q=0,-\sigma}
c_{q=0,-\sigma}| GS \rangle$.  Fig.\,4 shows a rapid change-over
when we sweep $\alpha$ through a narrow regime around $\alpha_c =
2.12$ for $\omega_0/t = 0.5$. It is in this narrow regime that we
can expect resonant pairing, which means a tunneling of
bipolaronic hole pairs from the central site to the plaquette
sites and vice-versa, in this way inducing a hole pairing on the
latter.

The efficiency for double charge fluctuations is controlled by the
ease with which holes tunnel in and out of sites. At zero
temperature it is given by the ground-state static correlator
\begin{eqnarray}
F^{\rm el}_{\rm exch} = \langle GS|c^{\dagger}_{q=0,\sigma}c_{5,\sigma}|GS \rangle
\end{eqnarray}
where $|GS\rangle$ denotes the ground state for the two-hole
system. The variation of $F^{\rm el}_{\rm exch}$ with $\alpha$ is
illustrated in Fig.\,5. We notice that for $\alpha$ close to zero,
the efficiency for such tunneling is governed by the residual
tunneling rate determined by the difference in energy between the
central site and the lower-lying Bloch states on the plaquette,
equal to $\Delta-2t$. As one approaches a certain $\alpha_c$ from
below, we notice a strong increase in this tunneling efficiency.
However, when going above $\alpha_c$, this efficiency drops
rapidly to zero, meaning that the holes are now quasi-localized on
the polaronic site in form of bipolarons.

In order to have a double charge fluctuation-induced
superconducting phase, driven by dynamical local lattice
instabilities, the above criterion for a high efficiency rate of
hole tunneling in and out of polaronic sites is not sufficient.
What is relevant is the efficiency for coherent double charge
fluctuations between the polaronic site and the plaquette. Such
coherent, or resonant, pair tunneling is given by the efficiency
factor
\begin{eqnarray}
F^{\rm pair}_{\rm exch} = \langle
GS|c^{\dagger}_{5,\uparrow}c^{\dagger}_{5,\downarrow}
c_{q=0,\downarrow}c_{q=0,\uparrow}| GS \rangle - [F^{\rm el}_{\rm
exch}]^2,
\end{eqnarray}
where we have subtracted out the rate of exchange of the bipolaron
with two holes on the plaquette via incoherent single-hole
processes, given by the second term in this expression. The
efficiency factor for resonant pair tunneling (see Fig.\,6) is
sharply peaked near $\alpha_c$, its maximal intensity depending on
the adiabaticity ratio $\omega_0/t$. The coherent pair tunneling
represents only a small fraction of  the total tunneling rate,
determined largely by its incoherent part. Nevertheless, the size
of this coherent component of the two-particle spectral function
is sufficient to ensure a high enough condensate fraction in order
for a superconducting phase-correlated state to be stabilized. It
provides a strong support in favor for such a mechanism to play a
role in real materials which have tendencies to local dynamical
lattice instabilities driving double valence fluctuations.
\begin{figure}
\begin{center}
    \includegraphics*[width=3in,angle=0]{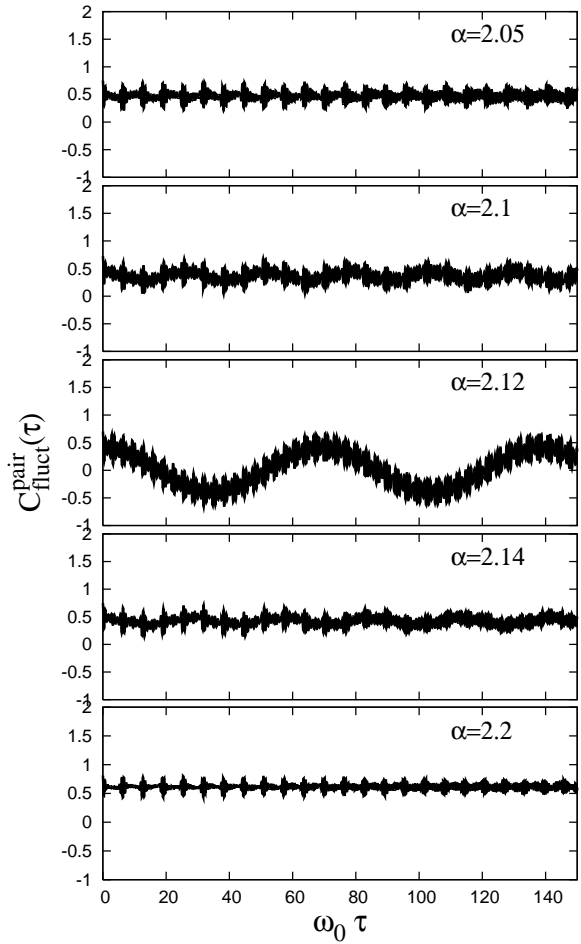}
\caption{\footnotesize Evolution in time $\tau$ of the resonant
tunneling of a hole pair in and out of a polaronic site, composed
of a cation together with its surrounding ligand environment, for
$\omega_0/t = 0.5$.}
\end{center}
\end{figure}
\begin{figure}
 \begin{center}
    \includegraphics*[width=3in,angle=0]{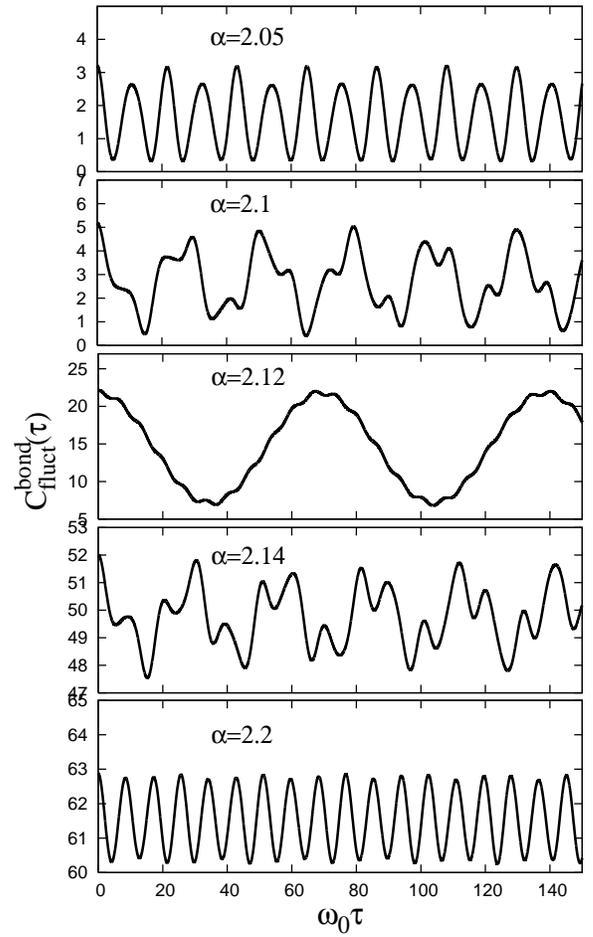}
\caption{\footnotesize Evolution in time $\tau$ of the bond-length
of the ligand environment which accompanies the resonant tunneling
of a hole pair in and out of the cation-ligand complex, for
$\omega_0/t = 0.5$.}
\end{center}
\end{figure}

In order to visualize resonant pair tunneling in and out of the
polaronic cation-ligand complex, we examine the evolution with
time $\tau$ of double charge fluctuations (see Fig.\,7). It is
described by the correlation function
\begin{eqnarray}
\!\!\!\!\!\!\!\!C^{\rm pair}_{\rm fluct}(\tau)& =& \nonumber \\
&&\!\!\!\!\!\!\!\!\!\!\!\!\!\!\!\!\!\!\!\!\!\!\!\!\langle GS
|[n_{k=0,\uparrow}(\tau)n_{k=0,\downarrow}(\tau)-
n_{5,\uparrow}(\tau)n_{5,\downarrow}(\tau)]\nonumber\\
&&\!\!\!\!\!\!\!\!\!\!\!\!\!\!\!\!\!\!\!\!\!\![n_{k=0,\uparrow}(0)
n_{k=0,\downarrow}(0)-n_{5,\uparrow}(0)
n_{5,\downarrow}(0)]GS \rangle.
\end{eqnarray}
Near $\alpha_c = 2.12$ (for $\omega_0/t = 0.5$) the charge
disproportionation follows a smooth oscillation in time with a
period $\omega_0 \tau \simeq 70$ and an amplitude varying between
$+0.5$ and $-0.5$. When this correlation function is positive and
at its maximum, such as for $\omega_0 \tau = 0, 70, ...$, it
indicates that both holes are on the central polaronic site. If it
is negative and maximal, such as for $\omega_0\tau = 35, 105,
...$, it means that both holes are on the plaquette. The tunneling
frequency of the coherent part of this motion of the hole pairs is
$\omega_0^* = [2 \pi / \omega_0\tau]\omega_0 \simeq [2 \pi
/70]\omega_0 \simeq 0.09 \omega_0$, as we see from Fig.\,7. It
represents the frequency for a correlated mode, locking together
the dynamics of the charge and deformations of the ligand
environment, or, in our simplified description, of some
bond-length $X=\sqrt{\hbar/2M\omega_0}\,[a^{\dagger} + a]$. The
fluctuations of the latter are described by the correlator

\begin{equation}
C^{\rm bond}_{\rm fluct}(\tau) =  \langle GS | [a^{\dagger}(\tau)
+ a(\tau)][a^{\dagger}(0) + a(0)]|GS \rangle
\end{equation}
and its variation in time $\tau$ is illustrated in Fig.\,8. In
line with the fluctuations of the charge distribution, this
bond-length fluctuation shows for $\alpha \simeq \alpha_c$ the
same smooth oscillating behavior with frequency $\omega^*_0=
0.09$, centered around some mean deformation $\langle GS
|[a^{\dagger} + a]| GS \rangle$. The important feature of such
resonant pair tunneling is that the fluctuations, which accompany
the systematic motion of the charge distribution as well as of the
bond-length,

(i) are on a time scale, typically of the order of $1 / \omega_0$,
much shorter than the one of the coherent tunneling motion $\sim
1/\omega^*_0$;

(ii) have an amplitude which is small compared to the amplitude of
the systematic coherent oscillating behavior on the time scale of
$1/\omega^*_0$.

As soon as we move away from this resonance regime, by tuning
$\alpha$ away from $\alpha_c$ for a given adiabaticity ratio, the
bond-length fluctuations very quickly lose their smooth coherent
oscillating behavior. They begin to show phase slips, which, as a
result, decorrelate the hole pair fluctuations from those of the
bond-length ones. As a result, the hole pairs get $confined$ on
either the central polaronic site for $\alpha \geq 2.14$ or else
on the plaquette sites for $\alpha \leq 2.10$. In our small system
this confinement is inferred from the absence of any sign of
periodic motion of the charge and bond-length fluctuations between
those two different kinds of sites. This is true for the whole
regime of $\omega_0 \tau $ between zero up to at least $20000$.
For all intents and purposes, one can therefore regard such a
behavior either as a localization for $\alpha \geq \alpha_c$ of
tightly bound hole pairs on the polaronic site, or as a
confinement for $\alpha \leq \alpha_c$ of two uncorrelated holes
on the plaquette. Incorporating this feature into an infinite
lattice structure, such as the one depicted in Fig.\,1, we expect
that for $\alpha \geq \alpha_c$ we have a true charge
disproportionation, i.e., the system breaks up into a mixture of
cations with localized bipolaronic hole pairs on the cation-ligand
sites and leaving the remaining cations without any hole. For the
opposite case, $\alpha \leq \alpha_c$, we expect a metallic
behavior with locally correlated pairing of holes induced by a
dynamical charge exchange coupling with the neighboring polaronic
sites. Depending on the overall particle density, this can give
rise to: (i) charge-ordered states with translational symmetry
breaking, passing from a covalent to ionic
bonding,\cite{Goodenough-1994} (ii) correlation-driven insulating
states.\cite{Robin-1998,Cuoco-2006}

\begin{figure}
 \begin{center}
    \includegraphics*[width=3.0in,angle=0]{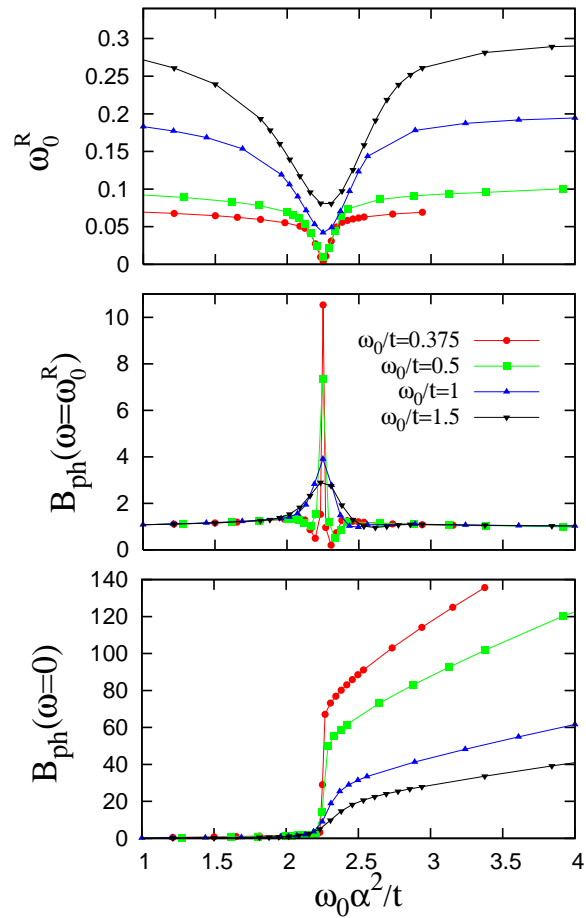}
\caption{\footnotesize (Color online) Lattice dynamical signatures
related to resonant bipolaronic double charge fluctuations for
several adiabaticity ratios $\omega_0/t$ and as a function of the
electron-lattice coupling $\omega_0\alpha^2/t$. Top panel shows
the softening of the intrinsic bond-length vibrational mode as one
approaches the critical coupling for resonance tunneling at
$\alpha_c \simeq \sqrt{(t +\Delta/2)/\omega_0}$, depending on
$\omega_0/t$. The maximal softening, occurring for
$\omega_0/t=0.375, 0.5, 1, 1.5$ at $\alpha_c = 2.45, 2.12, 1.5,
1.22$, respectively, is given by
$\omega_0^R(\alpha_c)/t\equiv\omega_0^*/t = 0.00336, 0.00907,
0.04205, 0.08103$. The central panel shows the associated
evolution of the spectral weight $B_{\rm ph}(\omega=\omega_0^R)$
of this renormalized mode. The bottom panel shows the spectral
weight $B_{\rm ph}(\omega=0)$ of the zero-frequency mode, related
to a static overall shift of the bond-length.}
\end{center}
\end{figure}

\section{Signatures of resonating bipolaronic double charge fluctuations}

The phase-correlated dynamics of the charge and the ligand
deformation results in several characteristic features in the
electronic and lattice properties which are in principle tractable
experimentally. Since these features arise from purely local
correlations, they are adequately described by the small cluster
model system studied here. These features involve:

(i) The strong softening of a local phonon mode with frequency
$\omega^R_0$ (see top panel of Fig.\,9) which describes correlated
fluctuations of the charge and the local lattice deformations of
the cation-ligand complex. This mode is described by the spectral
function for bond-length fluctuations, given by
\begin{equation}
B_{\rm ph}(\omega) = -\frac{1}{\pi} \, {\rm Im} D_{\rm ph}(\omega)
\quad\quad
\end{equation}
with
\begin{equation}
D_{\rm ph}(\omega) = \frac{2 M \omega_0}{\hbar}\, \langle\langle X
;X \rangle \rangle_\omega \quad .
\end{equation}
At finite temperature $B_{\rm ph}(\omega)$ is given by
\begin{equation}
B_{\rm ph}(\omega) = \frac{1}{Z}\sum_{n,m} e^{-\beta E_m}|\langle
m|(a^{\dagger} + a)|n\rangle|^2 \delta(\omega + E_n -E_m)
\end{equation}
with $Z = \sum_n e^{-\beta E_n}$. As this local mode softens upon
approaching $\alpha_c$, its spectral weight $B_{\rm
ph}(\omega=\omega^R_0)$ strongly increases, reaching a high
maximum at $\omega^R_0(\alpha_c) \equiv \omega^*_0$ (see middle
panel of Fig.\,9). This is indicative of the large amplitude
fluctuations of the bond-length which are necessary to drive the
double charge fluctuations between the polaronic cation-ligand
complex and the plaquette sites in a coherent fashion.

(ii) The spectral weight of the zero frequency bond-length
correlation function $B_{\rm ph}(\omega=0)$ is practically zero
for $\alpha \leq \alpha_c$, sharply increases in a narrow region
around $\alpha_c$, and then continues to grow linearly upon
further increasing $\alpha $ (see bottom panel of Fig.\,9). This
is indicative of a dynamical local lattice instability which
should manifest itself  as a {\it central peak structure} - a
quasi-elastic peak - in the neutron scattering cross-section. Such
features have been observed previously in systems  with local
dynamical lattice instabilities near ferro-electric phase
transitions,\cite{Cochran-1960} martensitic phase transition in
A15 compounds\cite{Shirane-1971} and
SrTiO$_3$.\cite{Coombs-1970,Riste-1973}

(iii) The emergence of a double-peak structure in the static Pair
Distribution Function (PDF) for the bond-length,
\begin{eqnarray}
g(X) &=& \frac{1}{Z}\sum_n e^{-\beta E_n} \langle n| \delta(x - X)
| n \rangle \nonumber \\ &=&  \frac{1}{Z}\sum_n e^{-\beta E_n}
\sum_k \sum_{\nu, \nu'} A^{(n)}_{k,\nu} A^{(n)}_{k,\nu'}\langle
\nu|\delta(x-X)|\nu' \rangle \nonumber \\ & \equiv & \sum_n
\,\gamma_n(X)
\end{eqnarray}
which can be measured by neutron spectroscopy or extended X-ray
absorption fine structure (EXAFS). This function has been
investigated in experimental studies concerning the local
dynamical lattice topology in the cuprate superconductors and in
the manganites, for the Cu$-$O and for the Mn$-$O bond stretch
mode, respectively (for a review see the contributions by Sinai
{\it et al.} and by Egami in Ref.\,25).

Our results, shown in Fig.\,10, indicate how the PDF evolves near
the critical value of $\alpha_c$, where resonant pairing is
present and how it is affected by the temperature. The left column
of this figure shows its evolution  as $\alpha$ crosses
$\alpha_c$, for several temperatures. At $T=0$ there is a clear
signature for a double-peak structure at $\alpha_c$, quickly
disappearing as we go away from this resonance condition. At very
low temperatures the double-peak structured PDF can be taken as an
indication for resonant pairing. At finite temperature, however,
one cannot accredit the double peak to a resonant tunneling of
double charge fluctuations. This becomes clear from the second and
third column of this figure, showing the contributions
$\gamma_0(X)$ and $\gamma_1(X)$ to the total PDF, coming from the
ground state and the first excited state, respectively. At finite
temperature, a double-peak structure always arises away from
$\alpha_c$ from thermal excitations which contribute to the PDF in
form of phase-uncorrelated single-peak contributions.

(iv) The onset of strong local pair correlations between the
charge carriers below a certain temperature $T^*$, resulting in a
pseudogap in the local density of states of our cluster. It was in
fact on the basis of the present scenario that a pseudogap phase
was initially predicted for the cuprates above the superconducting
transition.\cite{Ranninger-1988,Ranninger-1995} Its lattice-driven
origin became apparent in the isotope effect experimentally
observed\cite{Rubio-2000} on the temperature $T^*$ at which the
pseudogap opens up.\cite{Ranninger-2005} Within the same scenario,
the onset of the pseudogap feature was found to be related to a
change from single-particle transport at temperature above the
onset of the pseudogap ($T \geq T^*$) to one which is controlled
primarily by bound charge carrier pairs below $T^*$, upon
approaching $T_c$ from above. These bound pairs then acquire
free-particle-like dispersion, which results in a transient
Meissner effect\cite{Devillard-2000} (experimentally
verified\cite{Corson-1999}) and remnant Bogoliubov shadow modes in
the pseudogap phase.\cite{Domanski-2003} These theoretical results
were obtained by assuming an effective local static exchange
mechanism between bipolaronic bound hole pairs and itinerant
uncorrelated pairs of holes. The present work shows to what extent
such a heuristic scenario, investigated within the Boson-Fermion
Model, holds true and to what extent a time dependent relaxation
of the cation-ligand environment, trapping momentarily the charge
carriers in form of bipolarons, permits one to use such an
effective double-charge exchange mechanism.

In Fig.\,11 we illustrate separately the contributions to the DOS
coming from the holes on the plaquette and those coming from the
central polaronic site. The first ones involve the region near the
center of the DOS and show a significant amplitude as well as
asymmetry. That latter is a consequence of the asymmetric
charge-lattice deformation coupling which we assumed in this work
(see Section 2). The holes on the polaronic site are responsible
for the peaks on either side of the center of the DOS at large
frequencies. We notice that as the temperature decreases (right
hand panels of this figure) the pseudogap-like structure turns
into a well defined gap. It is this feature which, on the basis of
an effective Boson-Fermion Model
treatment,\cite{Cuoco-2006,Stauber-2007} finally leads to a true
superconducting gap below $T_c$.

(v) Resonant pair tunneling in and out of the polaronic
cation-ligand complex into the surrounding sites on the plaquette
incites pair correlations in the latter. They give rise to the
residual pairing in a many-body state, where local diamagnetic
fluctuations are embedded in an underlying Fermi sea and
eventually result in a macroscopic superconducting phase.
Indications for such local diamagnetism in the pseudogap phase of
the cuprates have been observed.\cite{Wang-2005} We illustrate in
Fig.\,12 the variation with $\alpha$ (near $\alpha_c$) of the
average density $n_p$
\begin{equation}
n_p = \langle n_{q=0,\uparrow}n_{q=0,\downarrow}\rangle - \langle
n_{q=0,\uparrow}\rangle \langle n_{q=0,\downarrow}\rangle
\end{equation}
of diamagnetic paired holes on the plaquette and notice that is it
quite sizeable - about $20\%$ of the total average density of
holes on the plaquette at $\alpha_c$. Slightly away from this
resonant regime, the holes are either pair-uncorrelated (for
$\alpha \leq \alpha_c$) or not existent on the plaquette (for
$\alpha \geq \alpha_c$), being confined to the polaronic
cation-ligand complex as localized bipolarons. The variation with
the temperature of $n_p$ is illustrated in Fig.\,13. The
contributions to $n_p$ come primarily from two sources. The first
and dominant one comes from the first excited state of the system,
lying an energy $\omega_0^*$ above the ground state, and reflects
the resonant nature of this diamagnetic pairing. The second
contribution comes from the ground state itself, which indicates a
static diamagnetic polarization of the holes induced by this
resonant pairing effect. As a result, as the temperature
increases, $n_p$ initially increases, then, it flattens off at a
certain temperature $T^*$, and eventually decreases to zero due to
thermal fluctuations.  We thus take $T^*$ as a measure of the
temperature for the onset of resonant pairing and the opening of
the pseudogap of the electronic DOS. The diamagnetism is strongest
in the adiabatic regime where the contribution coming from the
resonant pair tunneling is maximal, showing a peak of $n_p$ at low
temperatures. These resonant features are clearly reflected in the
dynamical pairing susceptibility (not presented here) which shows
a strongly peaked behavior at a frequency $\omega_0^*$ above the
ground state energy, indicative of the resonant nature of such
diamagnetic fluctuations.

\begin{figure}
 \begin{center}
    \includegraphics*[width=3.2in,angle=0]{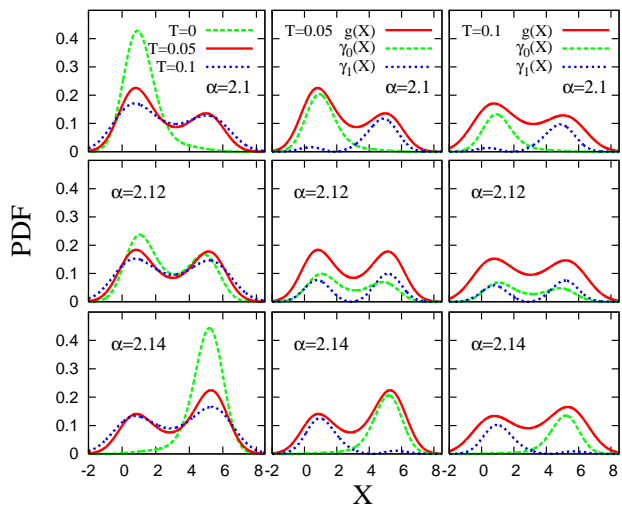}
\caption{\footnotesize (Color online) Left panels: pair
distribution function, $g(X)$, measuring the instantaneous
distribution of the cation-ligand bond-length at the central site,
for three values of $\alpha$ near $\alpha_c$ and three different
temperatures, for $\omega_0/t =0.5$. Central panels: pair
distribution function at $T=0.05$, together with the partial
contributions $\gamma_0(X)$ and $\gamma_1(X)$ coming from the
ground state and from the first excited state, respectively. Right
panels: same as in the central panels, for $T=0.1$.}
\end{center}
\end{figure}
\begin{figure}
 \begin{center}
    \includegraphics*[width=3in,angle=0]{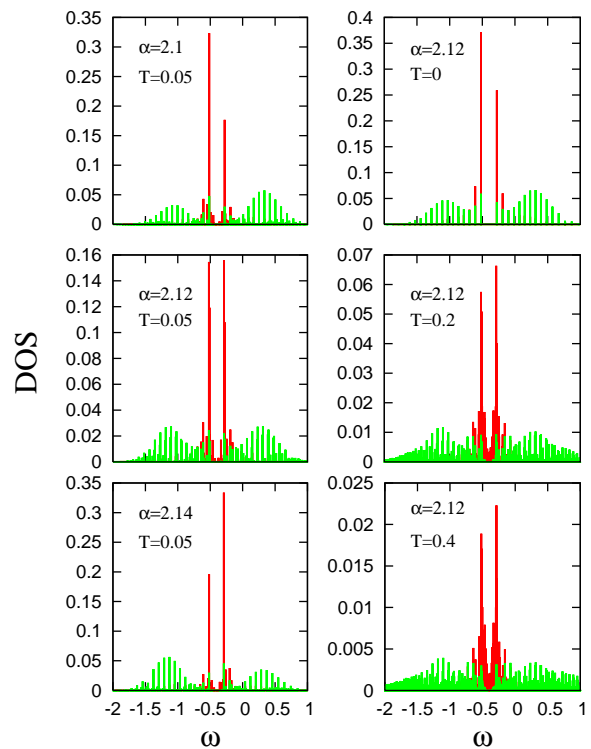}
\caption{\footnotesize (Color online) Single-electron DOS
decomposed into the contributions coming from the plaquette sites
(the spectral lines near the center of the DOS, in red) and from
the central polaronic site (the spectral lines at the wings at
high frequencies, in green). (a) Evolution of the DOS at low
temperature $T=0.05$ (left column) as $\alpha$ crosses the
critical value for resonant pairing at $\alpha = \alpha_c = 2.12$.
(b) Evolution of the DOS with increasing temperature at
$\alpha_c=2.12$ for $\omega_0/t=0.5$ (right column), which shows
the filling in of the pseudogap with increasing $T$. The energy
$\omega$ is in bare units and the pseudogap evolves around
$\Delta-2\hbar\omega_0\alpha_c^2\simeq -2t=-0.4$.}
\end{center}
\end{figure}
\begin{figure}
 \begin{center}
    \includegraphics*[width=3in,angle=0]{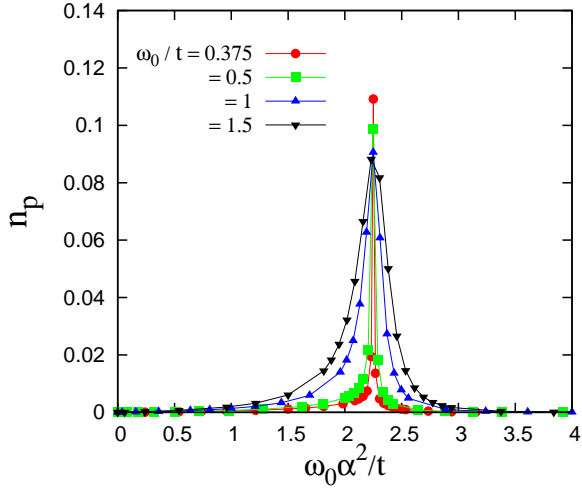}
\caption{\footnotesize (Color online) Density $n_p$ of correlated
hole pairs on the plaquette as a function of $\omega_0
\alpha^2/t$, showing a strong enhancement close to $\alpha_c
\simeq \sqrt{((t + \Delta/2)/\omega_0}$ due to resonant pair
tunneling. At $\alpha_c$, $n_p$ is typically of the order of
$0.1$, which has to be compared with the average total density of
holes on the plaquette, $\sum_{\sigma}\langle
GS|n_{q=0,\uparrow}n_{q=0,\downarrow}\rangle$, which from Fig.\,4
follows to be roughly equal to $0.5$.}
\end{center}
\end{figure}
\begin{figure}
 \begin{center}
    \includegraphics*[width=3in,angle=0]{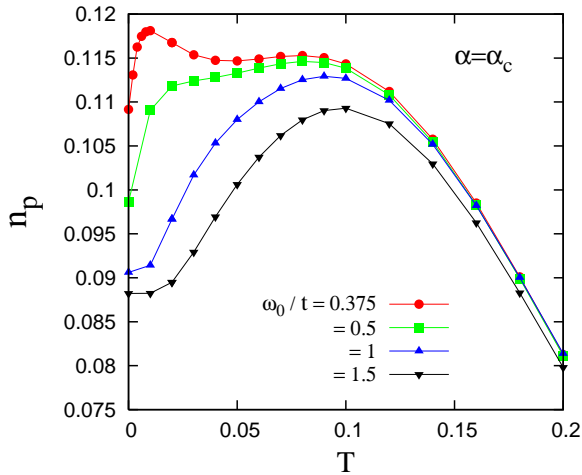}
\caption{\footnotesize (Color online) Variation with the
temperature $T$ of the density $n_p$ of correlated hole pairs on
the plaquette for various adiabaticity ratios $\omega_0/t$ and the
corresponding $\alpha_c$'s. With increasing $T$, $n_p$ rises and
then flattens off at some $T^*$, which indicates the onset
temperature for pair correlations on the plaquettes.}
\end{center}
\end{figure}

\section{Summary}

The main objective of the present work was to examine to what
extent double charge fluctuations, driven by local dynamical
lattice instabilities, can result in resonant pair tunneling in
and out of polaronic cation-ligand complexes, such that spatial
phase coherence of these fluctuations on a long-range scale can
ultimately be envisaged. This local physics with coherent resonant
pair tunneling is a prerequisite for ensuring the existence of
\underline{itinerant} bosonic quasi-particles describing
diamagnetic local pair fluctuations, embedded in a Fermi sea. It
results in the opening of a pseudogap in the single-particle DOS
at $T^*$, which above $T_c$ simulates that of a semiconducting
phase at finite $T$. As $T$ is decreased, this pseudogap goes over
smoothly into the superconducting gap below $T_c$ where part of it
is still controlled by the gap arising from pair correlations in
the normal state.\cite{Cuoco-2006} In the strong coupling
anti-adiabatic regime, resonant pair tunneling would reduce to an
exclusively bosonic system, composed of localized bipolarons
forming static charge disproportionation
insulators.\cite{Robin-1998} Superconductivity in systems with
resonant pairing is qualitatively different from that of a BCS
state in the sense that its $T_c$ is controlled by phase rather
than amplitude fluctuations, as is the case for the cuprate
high-$T_c$ superconductors in the underdoped regime. This
requires, on a dynamical basis, a correlated mixture of
diamagnetic pairs and quasi-free charge carriers, as described by
the Boson-Fermion Model in its simplest version, where Hubbard
correlation effects, which are certainly important in the
cuprates, are neglected.

Using the values of $\omega^*_0$ and $n_p$ deduced from the
numerical results illustrated in Figs.\,9 and 12, and referring to
the adiabaticity ratios $\omega_0/t = 0.375, 0.5, 1.0, 1.5$, we
estimate for the superconducting transition temperature the values
$T_c\simeq (\xi/a)[4, 10, 20,27]\,$K. Here we have used the
approximate expression $T_c \simeq \hbar^2 (\xi/a) (n_p/m_p\,
a^2)$ for phase fluctuation-driven superconductivity, where $\xi$
denotes the coherence length, $a$ the lattice constant, $n_p$ and
$m_p$ the number and the mass of the diamagnetic pairs. $m_p$ is
related to the resonant pair tunneling frequency $\omega_0^*$ by
$\hbar^2 / 2m_p = \omega^*_0 a^2$. Within our scenario of a
lattice instability-driven superconductivity, already such a crude
estimate gives values for $T_c$ which are quite reasonable. They
have been obtained by using a single input parameter, $\omega_0 =
50\,$meV, taken from experiments concerning the Cu$-$O bond
stretch mode. It is this mode which shows an anomalous softening
for large $q$-vectors close to the zone boundary and is thought to
play an active role in such possible dynamical local lattice
instabilities in the cuprate
superconductors.\cite{Tachiki-2003,Reznik-2006} The value of the
electron-phonon coupling $\alpha$ in this estimate for $T_c$ is
taken at the resonance, i.e., $\alpha= \alpha_c = [2.45, 2.12,
1.5,1.22]$ for the various values of the adiabaticity ratios
considered here. The nearest-neighbor and next-nearest-neighbor
hopping integrals are taken to be of the same order of magnitude,
i.e., $t/t^*=0.2/0.15$.

We can also estimate the onset temperature $T^*$ of pairing
correlations in the itinerant subsector of the charge carriers.
From the behavior of the pairing amplitude $n_p$, given in
Fig.\,13 as a function of the temperature $T$, $T^*$ should be
approximately identified by the crossing point of the tangent to
the rising portion of $n_p$ at low $T$ and the horizontal line
passing through the maximum of $n_p$. This $T^*$ is sensitive to
pressure which increases the local phonon frequency. Using the
thermodynamic relation between pressure-induced changes $\delta
T^*$ in $T^*$ and $\delta\omega^*_0$ in the local phonon frequency
$\omega^*_0$, $\delta T^*/T^* =\beta (\delta
\omega^*_0)/\omega^*_0$ ($\beta$ denoting a numerical factor of
order unity, related to the Gr\"uneisen parameter), this
experimentally verified relation\cite{Haefliger-2006} for the
cuprates with $\beta \simeq -3$ holds for our set of values for
$\omega_0^*$ and $T^*$ with a $\beta$ of order of unity.

Our concept of resonant pairing-induced diamagnetic fluctuations
bears some resemblance to what has been termed Cooper pairing due
to over-screening. Over-screening the Coulomb interaction between
the electrons implies a negative electronic contribution (besides
the positive ionic one) to the static dielectric constant,
$\varepsilon_{\rm el}(\omega=0, q) \leq 0$ for certain $q = k -
k'$ vectors in the Brillouin zone and provided that it is smaller
than $-1$. This over-screening does not result in static global
lattice instabilities.\cite{Kirzhnits-1976} It can thus provide a
mechanism for superconducting pairing among holes with wave
vectors $[k,-k']$ for much higher electron-phonon couplings than
initially thought. This idea has been brought up again in
connections with the cuprates superconductors, where the softening
of the Cu$-$O bond stretch mode in the Brillouin zone along the
directions of the anti-nodal wave-vectors, such as $[0,\pm
0.5]\pi/a$ and $[\pm 0.5,0]\pi/a$, was interpreted as a signature
for over-screening, resulting from local phase-correlated charge
and bond-length fluctuations.\cite{Tachiki-1989} As we could
confirm in the present work, such correlated pair fluctuations do
indeed exist and are quite robust. However, based on our study, we
do not expect that they give rise to a BCS-type superconducting
phase which is controlled by a pairing amplitude fluctuations.
Such systems should be close to an insulator-to-superconductor
transition,\cite{Cuoco-2004,Cuoco-2006} the insulator being
represented by phase-uncorrelated pair fluctuations and the
superconductor by spatially phase-correlated ones with a $T_c$
controlled by phase and not amplitude fluctuations. As far as
superconductivity is concerned, our approach to this
problem\cite{Stauber-2007,Cuoco-2006,Cuoco-2004,
Ranninger-1995,Ranninger-2005} differs from any standard
Eliashberg formulation.\cite{Tachiki-2003} In the Boson-Fermion
Model scenario superconductivity comes about via a center of mass
motion of dynamically fluctuating pairs. This is clearly
demonstrated by (i) the early Uemura muon spin resonance
experiments,\cite{Uemura-1989} showing a $T_c$ varying with the
concentration of superfluid carriers $n_p$, and (ii) a $T_c \simeq
2(\xi/a)[n_p \omega_0^*]$ which decreases with increasing
softening of the bond-stretch mode frequency\cite{Egami-2007}
$\omega_0-\omega_0^*$, where $\omega_0^*$ determines the inverse
mass of these itinerant diamagnetic pair fluctuations. This
experimental result\cite{Egami-2007} relates the doping dependence
of $T_c$ to the local lattice properties, characterized by a
softened local mode resulting from dynamical local lattice
instability for particular dopings in the superconducting regime.

This tempts us to make the following conjecture for our resonating
pair scenario, applied to the superconducting cuprates: Each
doping level in these superconducting compounds correspond to the
thermodynamically most unstable single-phase composition in the
synthesis. Once stabilized in their solid phase at low
temperature, the kinetic stability of those  metastable phases is
determined by our conditions for resonant pairing. This fixes, for
any given bare local mode with frequency  $\omega_0$ (relatively
insensitive to doping), a critical charge-ligand deformation
coupling $\alpha_c$, which will depend on doping and consequently
on the degree of softening of this mode.

Resonant pair tunneling poses rather stringent and experimentally
verifiable conditions on the local physics, as discussed here.
They are primarily related to features linked to dynamical lattice
instability driven by charge fluctuations, such as: (i) the strong
softening of a local phonon mode describing the correlated
charge-ligand deformation fluctuations (with a frequency
$\omega^*_0$ well below the bare local phonon frequency
$\omega_0$), (ii) a quasi-elastic peak in the neutron scattering
cross-section indicative of the dynamical nature of a local
lattice instability, (iii) a double-peaked pair distribution
function, accessible to neutron scattering spectroscopy and EXAFS,
(iv) locally fluctuating diamagnetism, and (v) a characteristic
structure of the electron spectral function, which shows polaronic
peaks at large frequencies on either side of the centre of the
pseudogap and a strongly peaked DOS near this center, coming from
the local diamagnetic pair correlations.

The natural next step of the present analysis will have to
incorporate these local dynamical features in a large scale
system. Here the question arises of the structure of the
superconducting state which not only will have to involve the
phase coherence of the charge but also of the lattice
instabilities driving the charge pairing. This implies cooperative
effects acting between such dynamically fluctuating local
clusters, which we expect to result in fluctuating topological
structures of the lattice, where insulating correlations compete
with superconducting ones.

A field theoretical formulation of the Boson-Fermion
Model,\cite{Cuoco-2004} not taking explicitly into account the
lattice degrees of freedom, has some similarity to the picture of
a coarse-grained superconducting state of spatially randomly
distributed bosonic modes.\cite{Balatski-2006} However, the main
difference between this approach and the one based on the
Boson-Fermion Model is that in the latter the local
inhomogeneities can fluctuate like in Josephson junction arrays
with inter-grain tunneling.

Treating the cuprates within this scenario will as a next step
require a microscopic treatment of the ligand environment. This
means: (i) explicitly taking into account  the different formal
valence states of the Cu-ligand complex, (ii) incorporating the
strong Hubbard U correlations in our present scheme, resulting in
local Zhang-Rice singlets, and (iii) study the dynamical charge
exchange between such cation-ligand complexes and the surrounding
metallic matrix beyond the single cluster model (discussed here)
and beyond any adiabatic approximation.\cite{Hozoi-2006} These and
other related questions will be dealt with in some future work.

\end{document}